\documentclass[preprint, times]{aastex631}

\usepackage{wrapfig}

\shorttitle{NASA DARES White Paper}
\begin{document}

\title{Habitable from the start: How initial planetary formation conditions may create habitable worlds}

\correspondingauthor{Benjamin Farcy}
\email{benjamin.j.farcy@nasa.gov}

\author[0000-0001-5159-6551]{Benjamin J. Farcy}
\affiliation{University of Maryland, Department of Astronomy, College Park MD 20742}
\affiliation{NASA Goddard Space Flight Center, Greenbelt MD 20771}

\author[0000-0002-0726-6480]{Darryl Z. Seligman}
\altaffiliation{D.Z.S. is supported by an NSF Astronomy and Astrophysics Postdoctoral Fellowship }
\affiliation{Dept. of Physics and Astronomy, Michigan State University, East Lansing, MI 48824, USA}

\author[0000-0001-8397-3315]{Kathleen E. Mandt}
\affiliation{NASA Goddard Space Flight Center, Greenbelt MD 20771}

\author[0000-0003-2152-6987]{John W. Noonan}
\affiliation{Physics Department\
Edmund C. Leach Science Center, Auburn University, Auburn, AL 36849, USA}

\author[0000-0002-9189-581X]{Sarah E. Anderson}
\affiliation{Aix-Marseille University, Laboratoire d'Astrophysique de Marseille, 13013, Marseille, France}

\vspace{-0.75cm}
\section{Introduction} \label{sec:intro}

The breadth of topics that encompass the search for life has expanded and evolved significantly since the emergence of the field of astrobiology. Initial astrobiology-centered investigations focused on detecting biosignatures in the Martian soil with the Viking lander. The field now encompasses identification of biosignatures throughout the galaxy and habitable worlds --- planets with sufficient liquid water and pre-biotic chemistry to support life. This evolution mirrors the improvement in our understanding of environments that may harbor life. The bulk planetary chemistry governs the habitability of a planet, which is in turn set  by the early solar system environment and planet formation processes. Therefore, investigations of solar and exoplanetary systems as a whole would provide insights into the factors that make a planet habitable.

%The field of astrobiology has evolved over recent decades, increasing the breadth and diversity of topics that encompass the search for life. Astrobiology first emerged in the Viking lander era when investigations focused on detecting biosignatures in the Martian soil; since then, the field has expanded to encompass searching for biosignatures and habitable worlds, or planets with sufficient liquid water and pre-biotic chemistry to support life. This approach is critical to the search for extraterrestrial life, as our understanding of past or currently extant environments that may harbor life has improved dramatically over the decades. However, the conditions that produce a habitable planet may be a consequence of bulk planetary chemistry, which itself is set and influenced by the early solar system environment and planet formation processes. Thus, a focus on planetary systems as a whole, including exoplanets and their early formation conditions, could inform on what produces a habitable planet. 

Bulk planetary chemistry govern planetary atmospheres, core sizes, magnetic fields,  heat engines, volatile inventories, and silicate mantle compositions. \textbf{We therefore advocate for investigations of formation conditions that establish planetary chemistry, and by extension, habitability.}

%Initial planetary chemistry and the formation environment of a planet in an early solar system have major controlling influences on the main components of a planet, such as: atmosphere, core size, magnetic field, planetary heat engine, volatile inventory, and silicate mantle composition. \textbf{To better aid in the search for habitable worlds the astrobiology community should work to understand the initial formation conditions that establish planetary chemistry, and by extension, habitability. 
%}

\section{Suggested Major Themes}
\subsection{Bulk Composition}
$\sim$93 wt.\% of the bulk composition of the terrestrial planets are Mg, Fe, Si, and O.  However, the \textit{ratios} of major elements (e.g. Mg/Fe, Si/O) differ between planetary bodies implying varying  mantle structures and mineralogies. These ratios govern  aspects such as  core radius, mantle composition/mineralogy, and felsic crust presence.  Prerequisites for the development of life (C, H, N, O, P, and S) also vary across planets.

% \begin{wrapfigure}[19]{L}{0.45\textwidth}
%     \centering
%     \includegraphics[width = 0.45\textwidth]{Figures/hypatia.pdf}
%     \caption{Major element abundances of bulk silicate planet (BSP) exoplanetary systems, estimated from metallicity observations of their host stars \citep{putirka2019composition}. Data is compiled as a part of the Hypatia catalog. }
%     \label{hypatia}
% \end{wrapfigure}

% Leftover nebular material from the formation of our solar system is commonly used to constrain the composition and processes in the protoplanetary nebula that led to the initial formation of the planets. Chondritic meteorites are understood as precursors to the planets, and the differences in the protoplanetary nebula are likely reflected in differences in chondrite composition. CI chondrites in particular match the composition of the solar photosphere remarkably well \citep{frank2023calcium}, demonstrating the link between host star metallicity and planetary building material. 
The bulk composition of a planet predominantly relies on the stellar metallicity which sets 
the major element composition of the protoplanetary nebula. Subsequent nebular processes that fractionate material between the inner and outer solar system may have driven the  diversity of  compositions of planets. Therefore, a focus on silicate-rich planets within an exoplanetary system may be critical to the search for life. Exoplanetary bulk composition may be derived from the  host star compositions \citep{putirka2019composition, spaargaren2020influence, unterborn2023nominal}. For example, stellar  metallicity in the Hypatia catalog \citep{hinkel2014stellar} are used as proxies of silicate mantle compositions for exoplanets.  Mass and radii constraints can also be used to infer the bulk density, major element composition, and mineralogy of an exoplanet.

%The bulk composition of a planet is, in part, established by the composition of its host star. Specifically, the metallicity of the star can determine the major element composition of the protoplanetary nebula, leading planetary systems to be reflective of their host stars. Subsequent nebular processes that fractionate material between the inner and outer solar system may have driven the diversity of planetary compositions we observe in our solar system. For exoplanetary systems, a focus on the bulk composition of the silicate portion of a group of planets within a stellar system may be critical to the search for life. Constraints on an exoplanet’s bulk composition may be derived from the composition of its host star \citep{putirka2019composition, spaargaren2020influence, unterborn2023nominal}. For example, the Hypatia catalog, a survey of the metallicity of $\sim$3,000 local stars \citep{hinkel2014stellar}, are seen as proxies of extrasolar systems and their silicate mantles. Additional constraints of exoplanet mass and radii to infer bulk density can be used to further understand the major element composition and mineralogy of a planet.

Exoplanetary atmospheric measurements can constrain  the bulk silicate composition. Specifically, short-period planets  experience sufficiently high surface temperatures for  silicate or oxide aerosols to contaminate their atmospheres \citep{gao2021aerosols}.  Atmospheric olivine and pyroxene (Mg$_{\text{2}}$SiO$_{\text{4}}$, MgSiO$_{\text{3}}$), or metallic oxides (CaTiO$_{\text{3}}$, Al$_{\text{2}}$O$_{\text{3}}$), should be detectable via transmission spectroscopy as aerosols on exoplanets with surface temperatures between 1,100 and 2,000 K \citep{wakeford2015transmission, parmentier2016transitions}.

% (CaTiO$_{\text{3}}$, Al$_{\text{2}}$O$_{\text{3}}$)
% (Mg$_{\text{2}}$SiO$_{\text{4}}$, MgSiO$_{\text{3}}$)

Future exoplanet survey missions such as the Habitable Worlds Observatory (HWO), along with geochemistry-based investigations of the rocky terrestrial planets in our solar system, can be used to observe the silicate portion of a group of planets as one large system. Comparing the terrestrial planets to each other in terms of their bulk composition and physical properties (e.g. atmosphere, core size, etc.), as well as to their host star, allows a much clearer understanding of what controls the potential habitability of a planet.\textbf{ We suggest that the astrobiology community adopt a comparative planetology approach to understanding exoplanetary systems, constraining bulk planet composition and mineralogy and host star metallicity as one complex and interconnected system.}
% \vspace{-0.45cm}
\subsection{Volatile Element Abundance} \label{vol_sect}
The abundance and distribution of planetary volatiles influences critical habitability components such as the mantle composition, atmosphere, core radius, and presence of surface liquid water. The volatility of an element is  the temperature at which 50\% of it condenses from a gas to solid  (T$_{50}$ at 10$^{-4}$ bar). The  life-forming elements (C, H, N, O, P, and S) have  low T$_{50}$ values \citep{lodders2003solar}.   The abundance of volatile elements therefore has a major impact on planetary habitability.

\begin{wrapfigure}[16]{R}{0.45\textwidth}
    \centering
    \includegraphics[width = 0.45\textwidth]{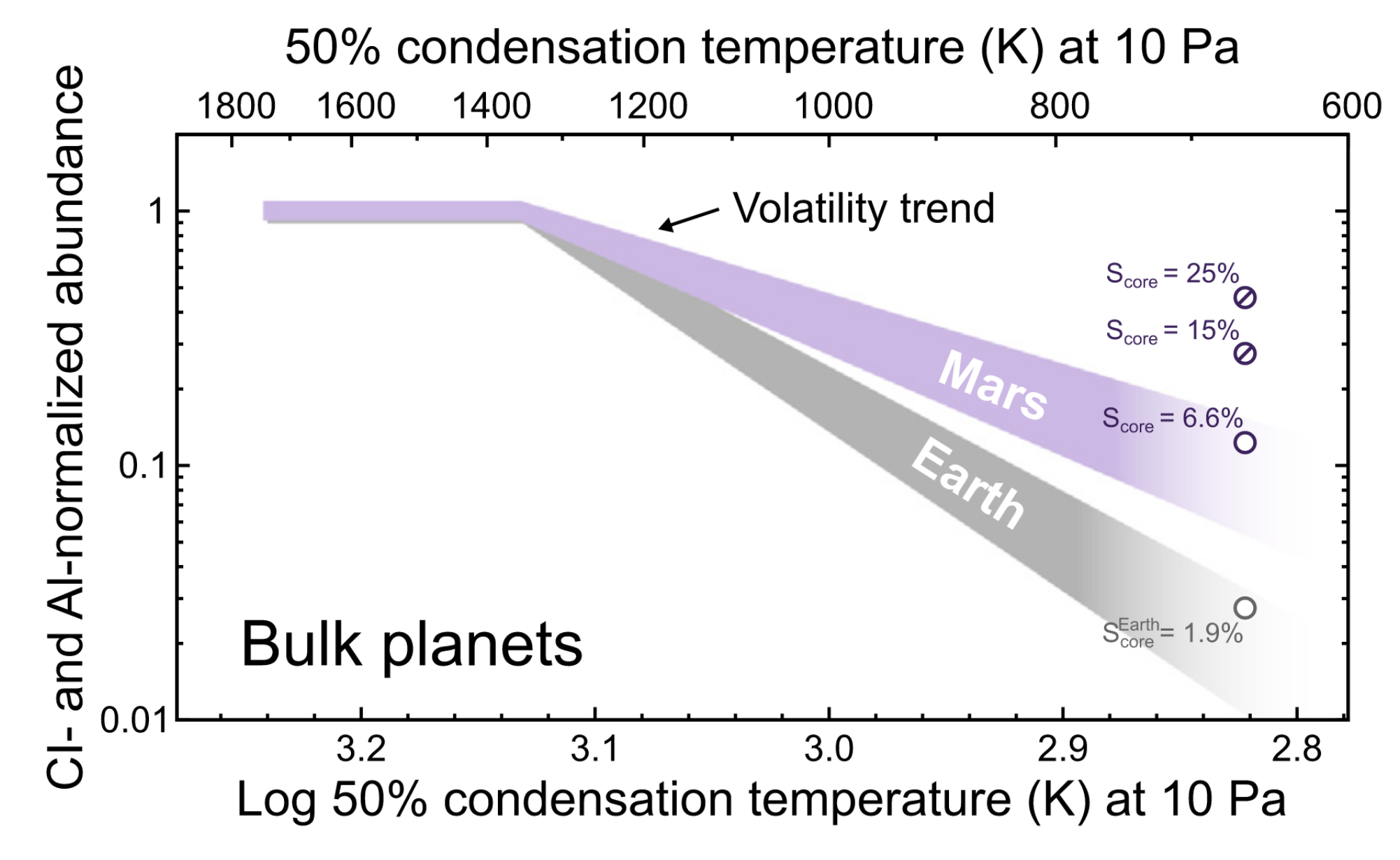}
    \caption{Volatile element abundance of bulk Mars and Earth relative to chondrites \citep{yoshizaki2020composition}. As volatile element abundance increases with heliocentric distance, Mars is enriched in volatiles. }
    \label{volatiles}
\end{wrapfigure}

The volatile element distribution throughout a planetary system drives the diversity of planetary composition, mineralogy, and structure. The abundance of volatiles in solar system planets increases with  heliocentric distance. The oxygen abundance in particular  controls  the silicate composition and core size. The oxygen fugacity (\textit{f}O$_2$), or partial pressure of O in a silicate system, determines mantle minerology and the elemental partitioning between the mantle, crust and core. For example, Mercury has low \textit{f}O$_2$, and therefore a predominantly olivine mantle (Si/O = 4) with high metallic Fe abundance, which leads to a large core \citep{nittler2017chemical}.  Mars is  more volatile-rich  (Fig. \ref{volatiles}, \cite{yoshizaki2020composition}) and contrastingly has a higher bulk \textit{f}O$_2$, a more enstatite-rich and FeO-rich mantle (Si/O = 3) and significantly smaller core.  It is worth note that the \textit{f}O$_2$ is independent of the stellar metallicity and dominated by protoplanetary nebula processes, which should be considered for habitability.

%In our solar system, volatile elements vary from one planet to the next, with increasing volatile abundance with increasing heliocentric distance from the sun. Oxygen in particular, one of the four major elements in the rocky inner solar system, has a major controlling influence on silicate composition and core size. The oxygen fugacity (\textit{f}O$_2$), or partial pressure of O in a silicate system, determines which minerals form in a planetary mantle, as well as the partitioning of elements into mantle and crustal minerals or planetary cores. Specifically, Mercury is characterized by a low \textit{f}O$_2$, producing a dominantly olivine mantle (Si/O = 4), and high metallic Fe abundance, which leads to a large core \citep{nittler2017chemical}. In contrast, Mars, a more volatile-rich planet \citep{yoshizaki2020composition}, has a higher bulk \textit{f}O$_2$, driving a more enstatite-rich mantle (Si/O = 3) and significantly smaller core with higher mantle FeO abundance. Thus, volatile element distribution throughout a planetary system drives the diversity of planetary composition, mineralogy, and structure. And while a planet's host star influences its composition, varying \textit{f}O$_2$ is an independent effect occurring in the protoplanetary nebula that should be considered for habitability.

% Fig. \ref{volatiles},

The presence of gas giants is a reflection of  early fractionation processes that move volatile elements to further heliocentric distances. Volatile-enriched planets form exterior to  condensation lines where a given volatile solidifies. High-volatility gas giants in our own solar system produced ocean worlds potentially capable of harboring life such as Europa \citep{ hand2009astrobiology} and Enceladus \citep{parkinson2008habitability}. %The habitability of other exoplanetary systems may be driven by a similar volatile distribution, as gas giants are also observed as exoplanets. \citep{reynolds1983habitability},

Post-formation material accretion can also deliver volatiles to planets. Isotopic ratios of volatiles such as C, N, H, and O have indicated that Earth’s volatiles may have been delivered from chondritic \citep{saal2013hydrogen, alexander2017origin} or cometary \citep{mandt2024nearly} sources. Prebiotic molecules, formed in a volatile-rich outer solar system, can also be delivered in large quantities to planetary surfaces \citep{chyba1992endogenous}. \textbf{ We suggest the astrobiology community works to better understand drivers of variations of volatile compositions across planets and the impact of volatile abundances on major planetary properties.}

%Additionally, while the initial nebular environment may influence volatile element distribution in a protoplanetary disk, later accretion of material can deliver volatiles after planetary accretion in our solar system and other exoplanetary systems as well. Isotopic ratios of volatiles such as C, N, H, and O have indicated that Earth’s volatiles may have been delivered from chondritic \citep{saal2013hydrogen, alexander2017origin} or cometary \citep{mandt2024nearly} sources. Prebiotic molecules, formed in a volatile-rich outer solar system, can also be delivered in large quantities to planetary surfaces \citep{ chyba1992endogenous}. 

%The volatile element content of a planet may be driven by early nebular or late accretion processes, and in turn, controls habitability by influencing mantle composition, core size, the presence of an atmosphere, and the abundance of life-forming elements. 

%\textbf{A concerted and unified effort by the astrobiology community to identify (i) the distribution, (ii) causes of varitions and (iii) the effect of abundance  on composition of volatile elements in the solar and exoplanetary systems would ___
\vspace{-11pt}

\subsection{Planetary Cores}

%The core of a planet is one of its three main reservoirs of material along with the  mantle and the crust. Properties of the core, including its size, composition, and solidification state are defining features of a planet. Moreover, these core features influence global planetary properties such as the presence  and strength of a planetary-scale magnetic field, mantle chemistry, and some aspects of the internal heating.

\begin{wrapfigure}[21]{L}{0.45\textwidth}
    \centering
    \includegraphics[width = 0.45\textwidth]{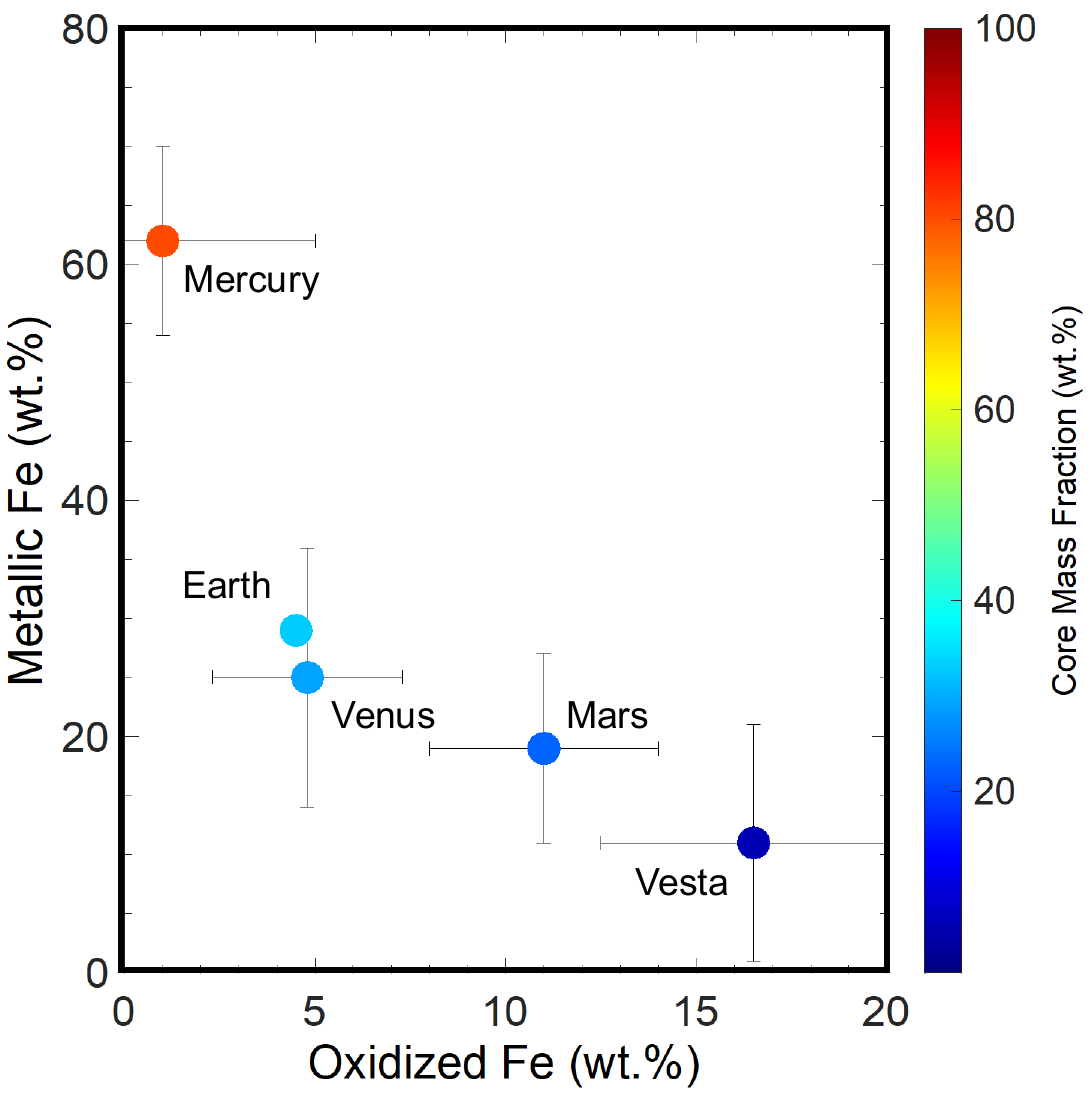}
    \caption{Ratio of metallic Fe to oxidized FeO of the rocky terrestrial planets, and its influence on the mass fraction of the planets core. Modified from \citep{cartier2019role}. }
    \label{cores}
\end{wrapfigure}

The size, composition, and solidification state  of a planetary core influences global planetary properties such as the presence  and strength of a planetary-scale magnetic field, mantle chemistry, and some aspects of the internal heating. The bulk O abundance determines the core radius (S\ref{vol_sect}). The \textit{f}O$_2$ of a planet determines the ratio of metallic Fe to oxidized FeO; mantles with lower \textit{f}O$_2$ and more metallic Fe produce larger cores.  The metallic Fe/FeO ratio increases with heliocentric distance in solar system planets because of their increasing volatile inventories.  In turn, more distant planets have smaller cores (Fig. \ref{cores}).   The core radius is therefore determined by the  initial local protoplanetary nebular conditions  which determine the volatile content. A better understanding of exoplanet systems and their relative core size contributions can inform on a potential gradient of formation of habitable planets in other stellar systems. In particular, high fidelity observations of planet mass and radius (and by extension, density), as well as computer modeling of the effect of stellar metallicity on core sizes, can provide valuable insights into exoplanetary cores.

% (Fig. \ref{cores})

The composition of the core is driven by  both initial nebular processes and the silicate mantle composition. Specifically, the \textit{f}O$_2$ of a silicate mantle determines which light elements will partition into a core during formation. S, Si, Mg, H, and O all partition into core material at varying \textit{f}O$_2$ states, which alters the core light element abundance. Variations in core composition will alter the viscosity, density, and thermal conductivity, all of which affect the geodynamo. Exoplanetary measurements of mass and radius can help constrain the bulk density of an exoplanet, which will provide insight into core density and light element abundance. 

The presence or absence of a planetary magnetic field is a result of a geodynamo driven by a convecting liquid outer core. This geodynamo is a direct consequence of the core size and composition. It is also determined by the  planet size which controls whether an outer core stays liquid or crystallizes. The magnetic field of the Earth is critical for its habitability because it shields the atmosphere from interaction with the solar wind and subsequent erosion. However, the geodynamo on Mars likely stopped at $\sim$3.8 Ga because of crystallization of the liquid outer core due to heat loss. This loss of the magnetic field enabled solar wind-driven atmospheric erosion and likely led to the removal of Martian surface water \citep{dehant2007planetary}. The observation of Earth-sized exoplanets with thick atmospheres imply the protection from a strong magnetic field, and protection from their stellar environment. 

\textbf{We recommend future investigations of (i)    the composition and size of exoplanetary cores and (ii) the presence of exoplanetary magnetic field. Observational constraints on planetary mass, radius, bulk density, magnetic field strength, and volatile element content can be used to understand exoplanetary core sizes and compositions. }

% All of these aspects are driven by initial protoplanetary nebula processes, meaning that initial core formation conditions may enable habitability in other solar systems. 
% \vspace{-0.5cm}
\subsection{Planetary Heat Engine}
Planetary heating drives   major planetary processes such as  plate tectonics, mantle convection, and volcanism. Interior planetary heat also degasses the planets mantle, moving life-forming elements (C, H, N, O, P, and S) from trapped deep mantle sources to accessible surface reservoirs such as atmospheres and oceans. A planet has two main sources of heat: residual heat from initial planetary accretion, and radiogenic heat from the decay of the heat-producing elements (HPEs), K, Th, and U. The relative amount of each of these two heat sources and the abundances of the HPEs depend on  formation processes within the early nebular environment and the stellar metallicity. For example, larger planets that have accreted more material have experienced more impact heating early in their history.

 Local stars exhibit a wide range in HPE abundances seen in the K/Th/U ratios and abundances \citep{unterborn2023nominal}, further driving internal heating within an exoplanet. These ratios  also vary in the solar system planets.   K is volatile while  Th and U are  refractory; therefore, the K/U and K/Th ratios are different for the Earth \citep{arevalo2009k, farcy2020k}, Mars \citep{yoshizaki2020composition}, the Moon \citep{prettyman2006elemental}, and chondrites \citep{lodders2003solar}. Moreover, the level of radiogenic heating of a given planet dissipates over time as more HPEs decay to stable decay products.

In addition to radiogenic and residual heat sources, tidal heating is likely a significant source of planetary heating. The eccentric orbit or orbital resonance of a planet can drive gravitational strain on a planetary mantle, inducing melting. Tidal heating drives volcanism on Io \citep{Peale1979,Smith1979,Morabito1979,Strom1979}.  Exoplanets on eccentric orbits could have orders of magnitude more tidal heating per unit mass than Io if they had the same quality factor \citep{seligman2024potential}. Conversely, residual heat loss is more efficient for smaller planets  due to less insulating mantle material and larger surface area to volume ratios.

The planetary heat engine is  critical for the emergence of life  because it determines the presence of (i) a convecting mantle, (ii) a liquid core and geodynamo, (iii) surface liquid water, and (iv) an atmosphere. Future observations of exoplanetary heating can help constrain planetary heat, and potentially, habitability. Examples of such observations include: the  presence or abundance of volcanic gases (SO$_2$, H$_2$S) and fine dust in transmission/reflectance observations, K, Th, and U metallicities of host starts, high-fidelity modeling of tidal heating, or other methods of understanding equilibrium temperatures. \textbf{We suggest that future exoplanet surveys for astrobiology should consider the role of planetary heat and its sources as a driver of habitability.}

\section{Conclusions}
The habitability of a rocky planet or ocean world depends on many critical factors that are often established in the early planetary formation environment. The ambient conditions of the protoplanetary nebula may govern planetary habitability  billions of years later.  The initial formation conditions are therefore  critical for the potential to host life. We argue that a community focus on comparative planetology will improve our understanding of what makes a planet habitable. Exoplanetary observations such as transmission or reflectance spectroscopy from the James Webb Space Telescope (JWST) or HWO, planetary masses and astrometry from the Kepler and Gaia telescopes, or higher fidelity models of planetary formation in different metallicity systems, can bolster our understanding of  planetary formation and  habitability. This approach should include a comparison of all of the terrestrial planets in a solar system relative to its host star, and to each other. Alternatively, it could compare the gas giants and small bodies relative to rocky terrestrial planets. We conlcude that a focus on early planetary formation processes would be a benefit to the Astrobiology community.

% Prioritizing comparative planetology would promote research within our solar system to establish the origin of reservoirs and processes that govern habitability for exoplanet systems.

% \section{References}

\section{acknowledgments}

D.Z.S. is supported by an NSF Astronomy and Astrophysics Postdoctoral Fellowship under award AST-2303553. This research award is partially funded by a generous gift of Charles Simonyi to the NSF Division of Astronomical Sciences. The award is made in recognition of significant contributions to Rubin Observatory’s Legacy Survey of Space and Time. This material is based upon work supported by NASA under award numbers NNH21ZDA001N-DALI and 80GSFC24M0006.

\bibliography{sample631}{}
\bibliographystyle{aasjournal}



\end{document}